\documentclass[9pt,twocolumn,twoside]{osajnl}

\journal{ol} 

\setboolean{shortarticle}{true}
\usepackage{siunitx}

\title{Generation and characterization of frequency tuneable sub-15~fs pulses in a gas-filled hollow-core fiber pumped by a Yb:KGW laser}
\author[1,*]{Mohammed Sabbah}
\author[1]{Federico Belli}
\author[1]{Christian Brahms}
\author[2,3]{Fei Yu}
\author[4]{Jonathan Knight}
\author[1]{John C. Travers}

\affil[1]{School of Engineering and Physical Sciences, Heriot-Watt University, Edinburgh, UK}
\affil[2]{Key Laboratory of Materials for High Power Laser, Shanghai Institute of Optics and Fine Mechanics, Chinese Academy of Sciences, Shanghai 201800, China}
\affil[3]{Hangzhou Institute for Advanced Study, University of Chinese Academy of Sciences, Hangzhou 310024, China}
\affil[4]{Centre for Photonics and Photonic Materials, Department of Physics, University of Bath, Claverton Down, Bath, BA2 7AY, UK}

\affil[*]{M.Sabbah@hw.ac.uk}

\begin{abstract}
We investigate soliton self-compression and photoionization effects in an argon-filled antiresonant hollow-core photonic crystal fiber pumped with a commercial Yb:KGW laser. Before the onset of photoionization, we demonstrate self-compression of our 220~fs pump laser to 13~fs in a single and compact stage. By using the plasma driven soliton self-frequency blueshift, we also demonstrate a tuneable source from 1030 to $\bold{\sim700}$~nm. We fully characterize the compressed pulses using sum-frequency generation time-domain ptychography, experimentally revealing the full time-frequency plasma-soliton dynamics in hollow-core fiber for the first time.
\end{abstract}

\setboolean{displaycopyright}{False}

\begin{document}

\maketitle
Hollow-core photonic crystal fibers (HC-PCFs) are a specific type of optical fiber with a micro-structured cladding and hollow core at the center \cite{russell_hollow-core_2014}. A particular type of HC-PCF, known as anti-resonant guiding fiber, has been used extensively to study the nonlinear interaction between laser light and gases. The long interaction length and broadband transmission windows of such fibers, in combination with the ability to tune the nonlinearity and dispersion through the gas type and pressure, have allowed their use in ultrafast nonlinear studies such as supercontinuum generation, ultraviolet resonant dispersive wave generation, light-plasma interaction, and optical soliton propagation \cite{travers_ultrafast_2011, russell_hollow-core_2014, markos_hybrid_2017}.

Pulse compression using soliton dynamics in HC-PCF has been studied extensively since the pioneering work by Ouzounov \textit{et al.}~\cite{ouzounov_generation_2003} in photonic-bandgap fibers, and Mak \textit{et al.}~\cite{mak_two_2013} in anti-resonant fibers~\cite{debord_multi-meter_2014, ermolov_carrier-envelope-phase-stable_2019, kottig_efficient_2020}. It is also a key component of many ultrafast dynamics in such fibers~\cite{joly_bright_2011, belli_vacuum-ultraviolet_2015}. Recently, pulse compression from \SI{340}{\fs} down to $\sim$ \SI{3.8}{\fs} was demonstrated in a two-stage experiment \cite{kottig_efficient_2020}. Subsequently, the same group also demonstrated pulse compression from \SI{250}{\fs} down to $\sim$ \SI{5.4}{\fs} at a repetition rate of up to \SI{10}{\mega\hertz} in a single stage assisted by chirped mirrors \cite{schade_scaling_2021}.

The tight fundamental-mode confinement of the light in combination with soliton self-compression creates pulses with a peak intensity well beyond the photoionization threshold of light gases and to the formation of plasma which can strongly act back on the light field. The polarizability of the free electrons is much larger and opposite in sign to the bound electrons \cite{boyd_nonlinear_2020, markos_hybrid_2017}, leading to a complex intensity-dependent (and hence time-dependent) drop of the refractive index which in turn causes an asymmetric phase modulation and a continuous shift of the spectrum towards higher frequency \cite{holzer_femtosecond_2011}. When combined with soliton dynamics, this effect leads to the soliton self-frequency blue-shift \cite{saleh_theory_2011}, which has also been observed in semiconductors \cite{husko_soliton_2013}. Recently, Huang \textit{et al.}~studied the soliton-plasma interaction in HC-PCF in detail \cite{huang_wavelength-tunable_2018, huang_continuously_2019, huang_ionization-induced_2019}. In Ref.~\cite{huang_ionization-induced_2019}, they demonstrated ionization-induced adiabatic soliton compression in HC-PCF starting with \SI{20}{\fs} pump pulses at \SI{800}{\nm}; similar dynamics were predicted in Ref.\cite{chang_combined_2013}. Initial experimental time-frequency characterisation of plasma-induced blue-shift during self-compression was carried out in the context of driving high-harmonic generation~\cite{tani_2017}. However, combined time-frequency measurements of clearly identifiable plasma-induced blue-shifting solitons in gas-filled hollow fibers have not previously been reported.

Here, we experimentally demonstrate the compression of \si{\micro\joule}-level pulses directly from a commercial \SI{220}{\fs} full-width half-maximum (FWHM) Yb:KGW laser to $\sim$ \SI{13}{\fs} in a single stage without the need for chirped mirrors. In addition, we demonstrate wavelength tunable sub-\SI{15}{\fs} pulses through soliton compression and soliton-plasma interactions in the same system. We fully characterize the temporally compressed pulses using sum-frequency generation (SFG) time-domain ptychography (TDP) \cite{witting_time-domain_2016}, and show that \SI{13}{\fs} can be obtained at \SI{1030}{\nm} using soliton compression. Moreover, we characterize the blue-shifting solitons in the time-frequency domain and show that they can compress down to \SI{15}{\fs}, and be tuned from the pump wavelength down to \SI{700}{\nm} simply by adjusting the input pulse energy. These measurements also provide the first unambiguous time-frequency measurement of the soliton self-frequency blue-shift.

\begin{figure}[t!]
\centering
\includegraphics[width=1.0\linewidth]{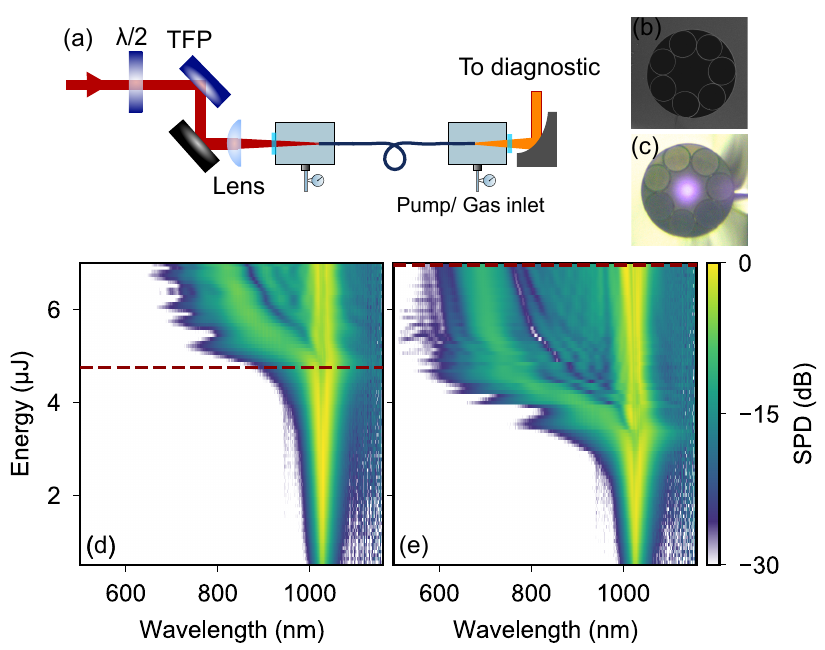}
\caption{(a) Optical layout of the experiment. $\lambda$/2: half wave-plate, TFP: thin-film polarizer. (b) and (c) Show the cross-section of the PCF used in the experiment and the near-field of the beam in a micrograph image for the end face of the fiber. (d) and (e) Experimental spectrum evolution for increasing coupled input energy, for \SI{220}{\fs} pulses launched into a \SI{29}{\micro\meter} core diameter, Ar-filled fiber with an increasing pressure gradient from vacuum to (d) \SI{800}{\milli\bar} and (e) \SI{1.2}{\bar}. The red dashed lines indicate the energy used for the time-domain characterization measurements.}
\label{fig:Setup}
\end{figure}

Figure~\ref{fig:Setup}(a) shows the experimental setup. A \SI{1030}{\nm}, \SI{220}{\fs} FWHM pump laser (Light Conversion PHAROS) was used at 1~kHz. Results at higher repetition rate, up to 200~kHz, are shown later. The pulse energy was controlled using a half-wave plate and a thin-film polarizer. The beam was then coupled into a \SI{29}{\micro\meter} core diameter single-ring anti-resonant PCF through a \SI{5}{\cm} focal length AR-coated plano-convex lens. The fiber is designed to suppress higher-order modes and maintain a clean fundamental Gaussian-like mode profile. The fiber, which had a core-wall thickness of $\sim$\SI{270}{nm}, was measured to have a broadband transmission window ranging from around \SI{630}{\nm} up to \SI{1370}{\nm} with less than 1~dB/m loss across that whole range, and around 0.1~dB/m at \SI{1030}{\nm}. The \SI{2}{\meter}-long fiber was sealed into the gas cells, so that a pressure gradient along the fiber could be created. Optical access was provided by an AR-coated \SI{2}{\mm} thick MgF\textsubscript{2} input window and uncoated \SI{1}{\mm} thick MgF\textsubscript{2} output window. We achieved up to \SI{76}{\percent} transmission at 1030~nm with both windows installed while the fiber was evacuated, this is less than expected from the linear loss, and we attribute the discrepancy to an imperfect focused spot size and launch conditions. The output spectrum was collected using an integrating sphere connected to a fiber-coupled CCD spectrometer. The spectrometer covers the spectral range \SI{200}-\SI{1160}{\nm} (Avantes ULS2048XL). The whole system is calibrated on an absolute scale with NIST traceable lamps.


Figure~\ref{fig:Setup}(d) shows the evolution of the experimental spectra for argon-filled HC-PCF with an increasing pressure gradient from vacuum at the input cell to \SI{0.8}{\bar} at the output cell. These parameters correspond to a soliton order of 5 at the output end of the fiber. The pulse spectrum broadens due to self-phase modulation which is accompanied by pulse compression in the time domain due to the effect of the anomalous group velocity dispersion (GVD). The pump wavelength lies in the anomalous GVD region throughout the fiber length. In addition, above \SI{5}{\micro\joule}, a strong spectral expansion towards the blue can be seen. This is due to the formation of plasma by the compressed pulse, leading to the soliton self-frequency blueshift, and is examined in more detail later.

\begin{figure}[t!]
\centering
\includegraphics[width=1.0\linewidth]{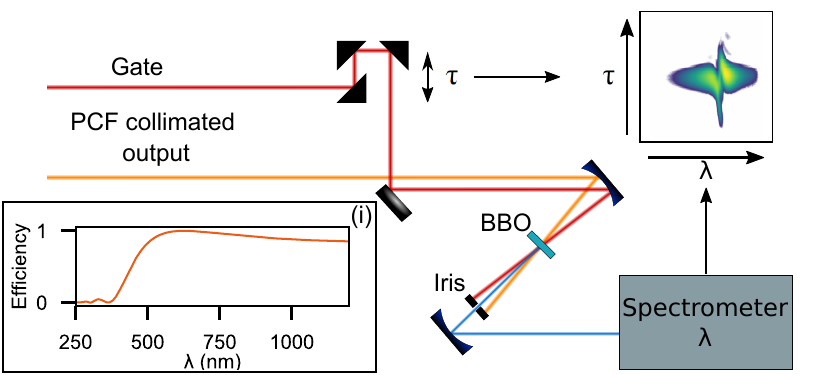}
\caption{(a) TDP characterization setup. The inset (i) shows the phase matching efficiency for SFG between the broadband test pulse and the narrowband gate pulse at \SI{1030}{\nm} using a \SI{10}{\micro\meter} BBO crystal for $\theta=\SI{29.2}{\degree}$.}
\label{fig:TDP}
\end{figure}

\begin{figure}[b!]
\centering
\includegraphics[width=1.0\linewidth]{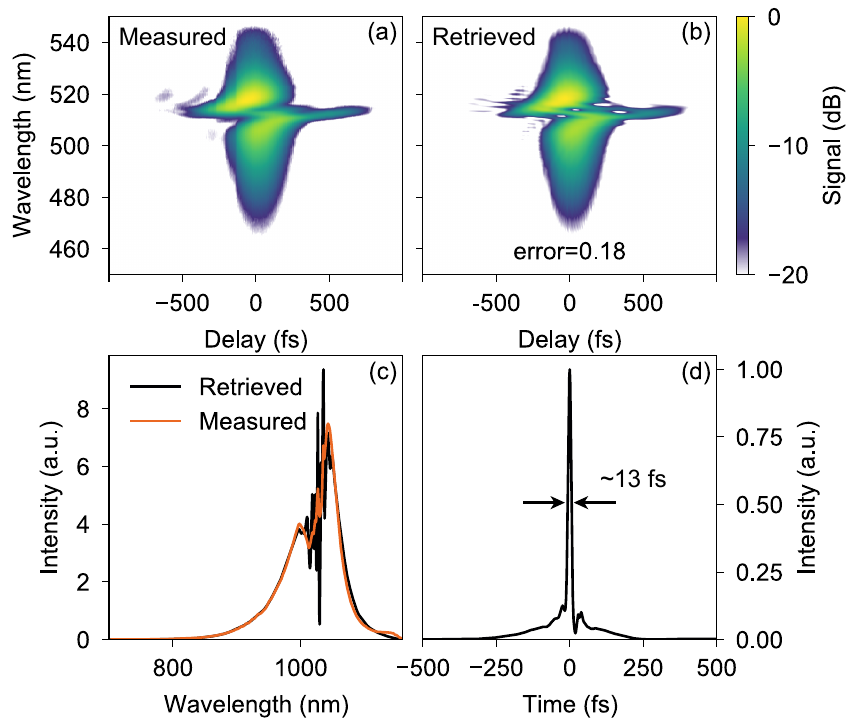}
\caption{(a) Measured and (b) retrieved SFG TDP traces on a logarithmic color scale for \SI{4.75}{\micro\joule} input pulses, using an increasing argon pressure gradient from vacuum to \SI{0.8}{\bar}. The RMS error between the measured and retrieved traces is $0.18\%$. (c) The measured and retrieved spectrum of the test pulse. (d) Temporal profile of the retrieved pulse from (b) at the crystal position. The FWHM duration of the pulse is \SI{13}{\fs}.}
\label{fig:SC_XFROG}
\end{figure}

We characterized the output pulses in the time domain using a home-built TDP setup shown in Figure.~\ref{fig:TDP}. We used the original laser pulse (the residual transmitted beam from the TFP) as the gate pulse. Both the PCF output (test) pulse and the gate pulse were focused into a BBO crystal in a non-collinear configuration. The BBO crystal was \SI{10}{\micro\meter} thick and cut for type-I (o-o-e) phase-matching ($\theta = \SI{29.2}{\degree}$), resulting in a phase-matching window that extends from around \SI{400}{\nm} to beyond \SI{1200}{\nm} as shown in inset (i) of Figure.~\ref{fig:TDP}. Spectral measurements for the test and gate pulses were made at the crystal position to obtain the frequency marginal for the TDP trace retrieval. We also measured the background noise of the TDP signal to help to get a cleaner spectrum. After the BBO, the signal was isolated from the test and gate beams by using a combination of Rochon prism and an aperture and then re-focused onto the spectrometer fiber tip. We used the extended ptychographic engine (ePIE) for pulse retrieval~\cite{maiden_further_2017}.

Figure~\ref{fig:SC_XFROG}(a) and (b) show the measured and the retrieved TDP traces for \SI{4.75}{\micro\joule} input pulses with a retrieval error of $0.18\%$. The overall retrieved pulse spectrum matches the measured spectrum well, as shown in Fig.~\ref{fig:SC_XFROG}(c). Note the asymmetry in the spectrum due to the self-steepening effect. The pulse FWHM at the fiber output is \SI{11}{\fs}---obtained by back-propagating the retrieved pulse numerically through the optics and air-path from the fiber output to the characterization crystal. The retrieved pulse duration at the crystal position, without back-propagation, is still short, with a duration of around \SI{13}{\fs}, as shown in Fig.~\ref{fig:SC_XFROG}(d). The main pulse is accompanied by a broad pedestal component, an inherent feature of soliton-effect pulse compression~\cite{pelusi_higher_1997}. The energy contained within the $1/e^2$ width of the compressed pulse is around 43\% of the total output pulse energy.

\begin{figure}[b!]
\centering
\includegraphics[width=1.0\linewidth]{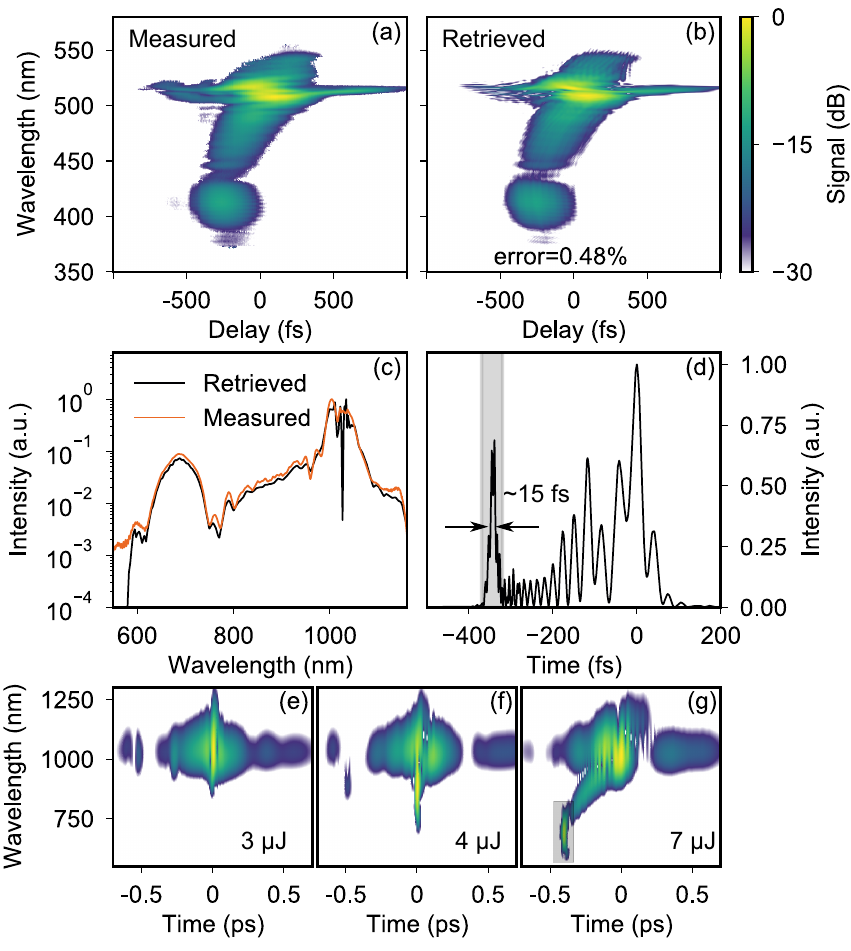}
\caption{(a) Measured and (b) retrieved  SFG TDP traces on a logarithmic color scale for \SI{4.75}{\micro\joule} input pulses using an increasing argon pressure gradient from \SI{0}{\bar} to \SI{1.2}{\bar}. The RMS error between the measured and retrieved traces is $0.48\%$. (c) The measured and retrieved spectrum of the test pulse. (d) Temporal profile of the retrieved pulse from (b). (e), (f), and (g) show the spectrogram evolution for the retrieved and back-propagated pulses at three different energies \SI{3}{\micro\joule}, \SI{4}{\micro\joule}, and \SI{7}{\micro\joule} respectively. The spectrograms are reconstructed from the retrieved and back-propagated pulses.}
\label{fig:blue_XFROG}
\end{figure}

Figure~\ref{fig:Setup}(e) shows the experimentally measured output spectrum as a function of pump energy when the output pressure is increased to \SI{1.2}{\bar}, increasing the nonlinearity. As the pulse propagates inside the fiber, it compresses from \SI{220}{\fs} to $\sim \SI{10}{\fs}$ and the spectrum broadens as shown. With further increase in the pulse energy, the intensity becomes sufficient to ionize the gas and form free electrons. The reduction in the refractive index due to liberated electrons is 1.8 times stronger than the refractive index change induced by the Kerr nonlinearity at \SI{4.25}{\micro\joule}~\cite{holzer_femtosecond_2011}. Hence, plasma dynamics are more pronounced and a blue-shifting pulse can be observed. Due to the significant anomalous dispersion experienced by the pulses, combined soliton-plasma dynamics occur, and the pump breaks up into self-frequency blue-shifting solitons~\cite{saleh_understanding_2011}. Above \SI{5}{\micro\joule}, the rate of the soliton blue-shift becomes slower, because ionization-induced loss decreases the pulse energy. By controlling the gas pressure and pulse energy, the central wavelength of the blue-shifting soliton can be easily tuned to the desired wavelength as shown in the two presented cases in Figure~\ref{fig:Setup}(d) and (e). We checked the stability of the blue-shifted soliton by measuring the relative intensity noise (RIN). We found that the RIN at the central wavelength of the blue-shifted soliton was 0.25\%, similar to the RIN of the pump laser system.

Fig.~\ref{fig:blue_XFROG}(a) and (b) show the measured and retrieved traces for \SI{7}{\micro\joule} pulse energy (indicated by the dashed line in Figure.~\ref{fig:Setup}(e)). We apply 40 iterations of the ePIE retrieval algorithm to retrieve the pulse profile from our measured traces with the spectral projection of the gate pulse applied for the first 20. The retrieved trace shows excellent agreement with the measured trace with an error of $0.48\%$. The retrieved pulse spectrum is shown in Figure~\ref{fig:blue_XFROG}(c) along with the separately measured spectrum. Again, an excellent agreement between the measured and retrieved spectrum is obtained. The retrieved pulse in the time domain is shown in Fig.~\ref{fig:blue_XFROG}(d). The blue-shifting soliton is located in the shaded area at the front of the pulse due to the blue-shift induced pulse acceleration (higher group velocity for higher frequencies in the anomalous dispersion region). The FWHM duration of the blue-shifting soliton is around \SI{15}{\fs} at the measured position. When back-propagated to the fiber end, the soliton duration is shorter, with a FWHM of \SI{13.5}{\fs}. The blue-shifting soliton contains \SI{0.5}{\micro\joule} of energy when the input pulse energy is \SI{7}{\micro\joule}. This gives a conversion efficiency of around 7\% (calculated as the ratio of blueshifting soliton energy to input energy). The evolution of the spectrograms shown in Figure~\ref{fig:blue_XFROG}(e--g) clearly confirms the solitonic nature of the blue-shifting pulse, as it remains parallel to the wavelength (frequency) axis, despite propagating over many dispersion lengths inside the fiber. This is the first unambiguous time-frequency measurement of the soliton self-frequency blue shift in hollow-core fibers.

\begin{figure}[t!]
\centering
\includegraphics[width=1.0\linewidth]{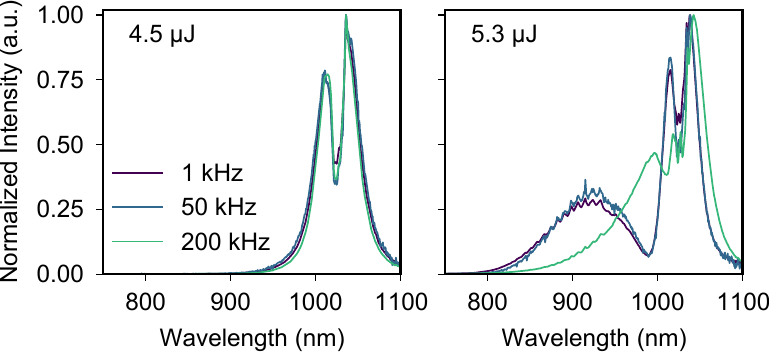}
\caption{Experimental output spectrum \SI{4.5}{\micro\joule} and \SI{5.3}{\micro\joule} at three different repetition rates.}
\label{fig:power}
\end{figure}

All results shown so far were obtained using 1 kHz repetition rate. We also investigated the behaviour of the plasma-soliton shaped spectrum as the repetition rate was increased. Fig.~\ref{fig:power} shows the output spectrum at different repetition rates for two energies and an increasing pressure gradient from vacuum  to \SI{0.8}{\bar}. For \SI{4.5}{\micro\joule}, the output spectrum is similar for the three repetition rates presented. At this energy, the peak intensity of the compressed pulse is not sufficient to ionize the gas and no inter-pulse effects can be seen in the spectra as the repetition rate is increased. On the other hand, when the pulse energy is increased to \SI{5.3}{\micro\joule}, an obvious difference in the output spectra can be seen between the \SI{200}{\kilo\hertz} case and lower repetition rates. We investigated this for different parameters at different repetition rates and found that when the pulse energy is high enough to start ionizing the gas inside the fiber, the spectrum at high repetition rates starts to deviate from lower repetition rates and its bandwidth decreases. This suggests that at high repetition rates, the plasma effects created inside the fiber by one pulse do not fully dissipate in between pulses and affect the following pulses, reducing or even inhibiting the blue-shift. Such an effect has been observed and studied in detail in Refs.~\cite{suresh_pump-probe_2019, koehler_post-recombination_2021, koehler_long-lived_2018}. In these articles, the authors suggested that plasma post-recombination heating and refractive index changes can strongly affect soliton compression starting from as low as \SI{50}{\kilo\hertz} repetition rate, as confirmed by our observations. Hence, care has to be taken to avoid plasma formation when developing a high-repetition-rate source, as described in Ref.~\cite{schade_scaling_2021}.

In summary, we experimentally demonstrate the flexible generation of short ultrafast laser pulses at different wavelengths from a commercial Yb laser in a single system based on gas-filled HC-PCF. Through soliton self-compression, we obtain pulses at \SI{1030}{\nm} with \SI{13}{\fs} duration on target without dispersion compensation after the fiber. By increasing the gas pressure, we enter the regime of soliton-plasma interactions and generate frequency-tuneable pulses down to \SI{700}{\nm} with \SI{15}{\fs} on-target duration through the soliton self-frequency blue-shift. We characterize the time-frequency structure of the frequency-tuneable compressed pulses using SFG TDP, and confirm the full soliton plasma dynamics in hollow-core fiber for the first time. Finally, we show that these dynamics can be scaled to 50~kHz in the presence of the plasma, but that at higher repetition rates inter-pulse plasma build-up inhibits the blue-shift. Our results enable simple sources of sub-\SI{15}{\fs} pulses using commercial and industrialised Yb-based ultrafast laser systems.

\begin{backmatter}
\bmsection{Funding}
This work was funded by the European Research Council under the European Union’s Horizon 2020 Research and Innovation program: Starting Grant agreement HISOL, No.~679649; Proof of Concept Grant agreement ULIGHT, No.~899900; and by the United Kingdom's Engineering and Physical Sciences Research Council: Grant agreement EP/T020903/1. F.B. acknowledges support from the Royal Academy of Engineering through Research Fellowship No. RF/202021/20/310. C.B. acknowledges support from the Royal Academy of Engineering through Research Fellowship No. RF/202122/21/133.


\bmsection{Disclosures} The authors declare no conflicts of interest.

\bmsection{Data availability}
The data that support the findings of this study are available from the corresponding author upon reasonable request.

\end{backmatter}

\bibliography{sample}

\bibliographyfullrefs{sample}


\ifthenelse{\equal{\journalref}{aop}}{%
\section*{Author Biographies}
\begingroup
\setlength\intextsep{0pt}
\begin{minipage}[t][6.3cm][t]{1.0\textwidth} 
  \begin{wrapfigure}{L}{0.25\textwidth}
    \includegraphics[width=0.25\textwidth]{john_smith.eps}
  \end{wrapfigure}
  \noindent
  {\bfseries John Smith} received his BSc (Mathematics) in 2000 from The University of Maryland. His research interests include lasers and optics.
\end{minipage}
\begin{minipage}{1.0\textwidth}
  \begin{wrapfigure}{L}{0.25\textwidth}
    \includegraphics[width=0.25\textwidth]{alice_smith.eps}
  \end{wrapfigure}
  \noindent
  {\bfseries Alice Smith} also received her BSc (Mathematics) in 2000 from The University of Maryland. Her research interests also include lasers and optics.
\end{minipage}
\endgroup
}{}

\end{document}